# Electronic band structure of epitaxial PbTe (111) thin films observed by angle-resolved photoemission spectroscopy

Zhenyu Ye[1], Shengtao Cui[2], Tianyu Shu[1], Songsong Ma[1], Yang Liu[1], Zhe Sun[3], Jun-Wei Luo[4], Huizhen Wu[1,*]

[1] Department of Physics, State Key Laboratory for Silicon Materials, Zhejiang University, Hangzhou 310027, China
[2] School of Physical Science and Technology, ShanghaiTech University and CAS-Shanghai Science Research Center, Shanghai 201203, China
[3] National Synchrotron Radiation Laboratory, University of Science and Technology of China, Hefei 230029, China
[4] State Key Laboratory of Superlattices and Microstructures, Institute of Semiconductors, Chinese Academy of Sciences, P. O. Box 912, Beijing 100083, China

* Email: hzwu@zju.edu.cn

**Abstract** Using angle-resolved photoemission spectroscopy (ARPES), we studied bulk and surface electronic band structures of narrow-gap semiconductor lead telluride (PbTe) thin films grown by molecular beam epitaxy both perpendicular and parallel to the Γ-L direction. The comparison of ARPES data with the first-principles calculation reveals the details of band structures, orbital characters, spin-orbit splitting energies, and surface states. The photon-energy-dependent spectra show the bulk character. Both the L and Σ valence bands are observed and their energy difference is determined. The spin-orbit splitting energies at L and Γ points are 0.62 eV and 0.88 eV, respectively. The surface states below and close to the valence band maximum are identified. The valence bands are composed of a mixture of Pb 6s and Te $5p_z$ orbitals with dominant in-plane even parity, which is attributed to the layered distortion in the vicinity of PbTe (111) surface. These findings provide insights into PbTe fundamental properties and shall benefit relevant thermoelectric and optoelectronic applications.

## I. Introduction

Narrow gap lead telluride (PbTe) has attracted tremendous interest because of its unique properties and important applications. It has been widely used in thermoelectric and midinfrared optoelectronic devices[1, 2], such as thermoelectric power generation with the highest efficiency at high temperature[3, 4], infrared detectors[5], light emitting diodes[6], and lasers[7]. Particularly, significant progress has been made in the past decades to improve the thermoelectric figure of merit of PbTe-based thermoelectric devices, including the application of Tl doping[3], band convergence engineering[8], and multiscale hierarchical design[4]. PbTe also hosts fascinating physical properties. Recently, Fu *et al.* theoretically predicted that PbTe could become a topological crystalline insulator by applying pressure, strain or forming an alloy with SnTe[9]. A two-dimensional electron gas (2DEG) with high mobility has been discovered by Zhang *et al.* at the interface of rock-salt PbTe and II-VI zinc-blende CdTe, and the quantum oscillation measurements demonstrated the Dirac fermion nature of the 2DEG [10].

The band structures of PbTe are pivotal to the understanding of its outstanding performances. There have been a plethora of theoretical studies of energy band structure of PbTe. The calculations of electronic band structure of PbTe have been carried out by using $\mathbf{k}\cdot\mathbf{p}$ perturbation theory[11], the empirical pseudopotential method[12], tight binding model[13], and first-principles calculations involving hybrid functional[14] and quasi-particle *GW* approximations[15]. For the experimental studies of the electronic band structure of PbTe, however, only band gaps at some specific k points were deduced from the optical reflectivity[16] and electro-reflectivity[17] measurements. These measured results were interpreted in terms of Van Hove's singularities from transitions between bands at high symmetry points and lines, without explicit momentum resolution in the entire Brillouin zone (BZ). Grandke *et al.*[18] performed measurements on (100) surface of single crystals of lead chalcogenides using rare-gas resonance lines at 16.85 eV and 21.22 eV, which was one of the first angle-resolved photoemission spectroscopy (ARPES) studies of the semiconductor surfaces. They claimed that the indirect-transition model could be applied in lead chalcogenides with $O_h$ symmetry due to a larger number of critical points along Γ-X of the BZ, whereas the direct-transition model is more applicable for Ge and zinc-blende semiconductors[18, 19]. Hinkel *et al.* [20] also performed photoemission spectroscopy (PES) measurements using synchrotron radiation on (100) surface of PbTe bulk crystal. The electron emission normal to the surface was recorded and the dispersions along Γ-X were obtained, and the direct-transition model was adopted to interpret the data. Though the (100) surface has been well studied, there is no systematic study regarding the (111) surface to our best knowledge. This is because the (111) surface is a strong polar surface and is difficult to cleavage from bulk single crystal, contrary to the

(100) surface which is the natural cleavage plane for rock-salt single crystal. The band edge states of PbTe which occur at L point (direct band gap) is the most important for optical and transport properties of PbTe. To perform ARPES measurements around the L point, the (111) surface is the most suitable. The challenge can be resolved by performing ARPES characterizations of the PbTe (111) thin films grown by molecular beam epitaxy (MBE).

On the other hand, the band inverted SnTe and its alloy with PbTe have been considered to be the prototype material of topological crystalline insulator[21, 22]. The study of the normal narrow-gap PbTe might provide valuable information on the physics of SnTe. For example, Tanaka *et al.* have performed ARPES measurements on PbTe (001) surface[23]. Yan *et al.* have taken ARPES spectra of $Pb_{1-x}Sn_xTe$ (111) thin films using helium lamp[24]. Their results show that PbTe does not host evident topological protected gapless surface states and only states near band edge were observed. However, high-resolution band mapping of PbTe (111) over a large energy and momentum scale and the detailed comparison with theoretical calculation is still missing for the important (111) surface, the study of which may contribute to our understanding of the physics in these materials.

In this paper, using synchrotron-based high-resolution ARPES, we examine the electronic band dispersion of PbTe (111) thin films grown by MBE. The comparison between the experimental data and first-principles calculations helps us to identify the bulk and surface electronic band structures of PbTe (111) thin films.

**II. Methods**

*Molecular beam epitaxy.* High-quality single crystal thin films (~1μm) of PbTe were grown on freshly cleaved (111) oriented $BaF_2$ substrates by MBE with a base pressure better than $2\times10^{-10}$ Torr. [10, 25]. The substrates were prebaked for 30 min at 500℃ before the growth of PbTe films. 1 μm PbTe films were grown at a rate of 1 μm/h (a beam flux of $5\times10^{-7}$ Torr) and the substrate temperature was kept at T=250°C. To keep equilibrium stoichiometry, an additional Te effusion cell was used during growth. The growth surface was monitored by reflection high energy electron diffraction (RHEED). Because $TeO_2$ can easily form on the PbTe surface when the samples are exposed to air, a selenium capping layer about 200 nm was deposited on the top surface of PbTe films to protect them from oxidation during the transfer from the MBE chamber to the ARPES facility [26].

*ARPES measurements.* The ARPES experiments were carried out at the beam line 13U of the National Synchrotron Radiation Laboratory (NSRL) at Hefei, China, using a Scienta R4000 electron spectrometer. Angle resolution was 0.3° and the combined instrumental energy resolution was better than 20 meV. All samples were measured at 25 K under a vacuum better than $5\times10^{-11}$ mbar. The Se cappling layer of PbTe samples was removed by sputtering for 10 min with 500eV Ar$^+$ ions and annealed for 5 min at a temperature of 250°C.

*First-principles calculation.* The electronic structures and total energy were calculated using first-principles density functional theory as implemented in the VASP package[27]. The projector augmented wave method was used with a plane-wave expansion up to 400 eV. The modified Becke and Johnson exchange potential (termed TB-mBJ)[28] was adopted to improve the description of bulk band structures. The TB-mBJ approach remarkably improved the local density approximation for p-p like band gap and the error of band gap is expected to be small. Spin-orbit interaction was included in the self-consistent calculation following the approach of Kleinman and Bylander[29]. Γ-centered $8 \times 8 \times 8$ Monkhorst-Pack k-mesh was used for the Brillouin zone integration for the bulk. For the slab calculation, we used the supercell slab geometry with symmetric slabs with either cation or anion at both surfaces to avoid the polar field for the (111) crystal direction. The thickness of the vacuum region separating periodically repeated slabs was more than 20 Å and was tested up to 40 Å. We tested the convergence of our results up to slab thicknesses of 95 atoms. Γ-centered $7 \times 7 \times 1$ k-mesh was used and the atomic geometries was allowed to relax until the Hellmann-Feynman forces became smaller than 10 meV/Å within generalized gradient approximation[30].

## III. Results and discussion
### A. Crystal structure and Brillouin zone

The face centered cubic (FCC) Bravais lattice of rock-salt crystal structure PbTe consists of two atoms. A view along the [111] direction of the FCC lattice shows six-layer periodicity and polar surface nature, which consists of alternating cation and anion layers as shown in Figure 1(a). No reconstruction was observed for the PbTe (111) surface, as illustrated in the RHEED patterns during MBE growth (Figure 2(a)), which is consistent with previous results[31].

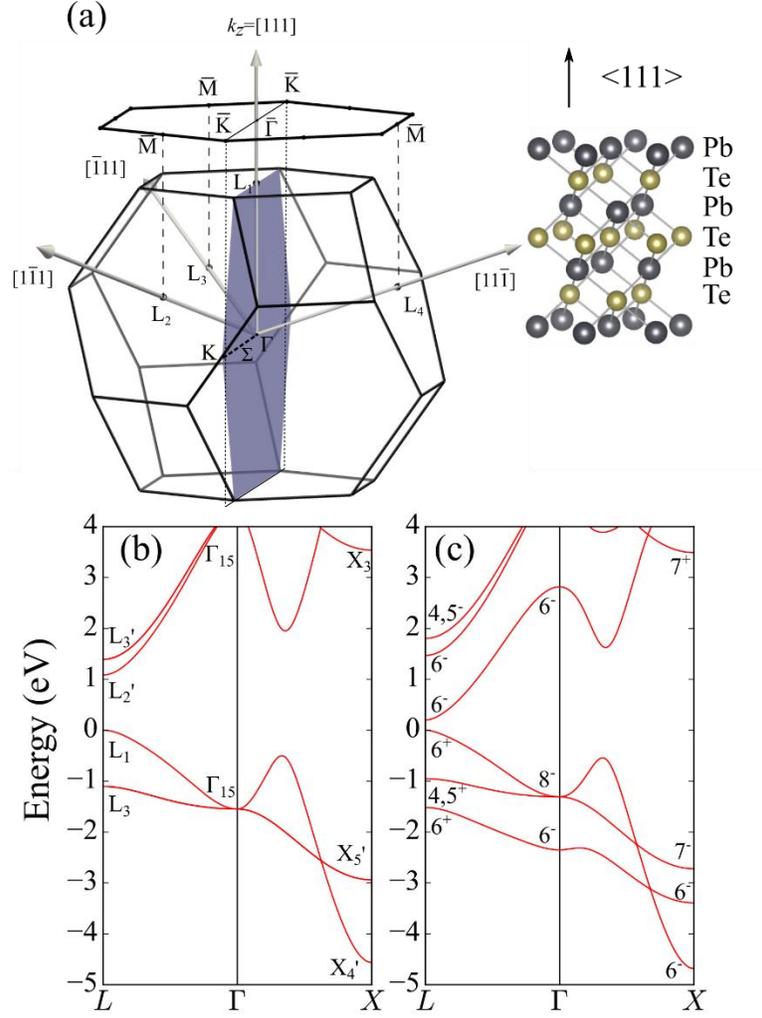

Figure 1. (a) Bulk FCC BZ and the surface BZ. Inset: Lattice structure of PbTe along <111> direction. Band structure of PbTe (b) without and (c) with spin-orbit coupling.

The direct band gap of PbTe is at four-fold valley degenerate L point of the FCC BZ[32], as shown in Figure 1(a). The surface Brillouin zone (SBZ) of PbTe is hexagonal. The band edge states at four L points are projected to one $\bar{\Gamma}$ point and three $\bar{M}$ points in hexagonal SBZ, respectively (Figure 1 (a)). The rock salt crystal structure of PbTe can be obtained by $\frac{1}{4}$ shift along [111] direction of the cation and anion sublattices from zinc-blende crystal structure. The additional inversion symmetry transforms $T_d$ point group of zinc-blende structure to $O_h$ point group symmetry of rock salt structure and makes the energy bands at eight time-reversal-invariant momenta (Γ, 4L, 3X) with definite parity, either positive parity or negative parity. The parity places strong constraint for the band coupling. The calculations of the band structure for PbTe without and with spin-orbit coupling are plotted in Figure 1(b,c). As shown in Figure 1(c), the lone pair Pb 6s $L_6^+$ band couples strongly with the Te 5p valence electrons at L point, pushing the $L_6^+$ valence band upwards, which makes the valence band maximum of PbTe at L point. On the other hand, the Pb p orbital couples strongly with high-lying conduction band with the same symmetry at L point to make the $L_6^-$ become conduction band minimum. The coupling is stronger at low temperature due to the shorter bond length. Thus PbTe has a direct band gap of 0.32 eV at room temperature and 0.19 eV at 4K at L point (Table 1).

**Table 1** Lattice constants, dielectric constants, effective masses, and energy band gaps of PbTe.

| $a$ (Å) (Ref. [14]) | ε(Ref. [33]) | $m^*/m_e$ (Ref [34]) | $E_g(L)$ (eV) | |
|---|---|---|---|---|
| | | | Calc. | Expt.(Ref. [35]) |
| 6.428 (30K) | $\varepsilon_0$=414, | $m_\perp^e = 0.022$, $m_\parallel^e = 0.185$ | 0.20 | 0.19 (4K) |
| 6.462 (300K) | $\varepsilon_\infty$=40 | $m_\perp^h = 0.025$, $m_\parallel^h = 0.236$ | | 0.32 (300K) |

**B. The $k_z$ Measurement**

The angle-integrated ultra-violet photoelectron spectroscopy measured with hν=91.5 eV photons shows the core energy levels of Pb and Te (Figure 2(a)), and the peak of Se is very weak after the Ar$^+$ ion bombardment. The inset in Figure 2(a) is a RHEED image for PbTe (111) growth surface by MBE. The streaky pattern indicates the smooth surface with layer-by-layer growth process, which is critical for the following ARPES experiment. To gain the dispersion of the band structure along k$_z$, we measured photon energy dependent photoemission spectra from 14 eV to 45 eV with a step of 1 eV. The dispersion of the first and the third valence band of L$_6^+$ symmetry are clearly observed as shown in Figure 2(b), proving their bulk origin. The top of valence band varies with the photon energy and the maximum of valence band energy is found to be at 19 eV. The comparison between photoemission spectra and first-principles calculations along Γ-L-Γ is shown in Fig. 2(c), and the agreement between them is evident.

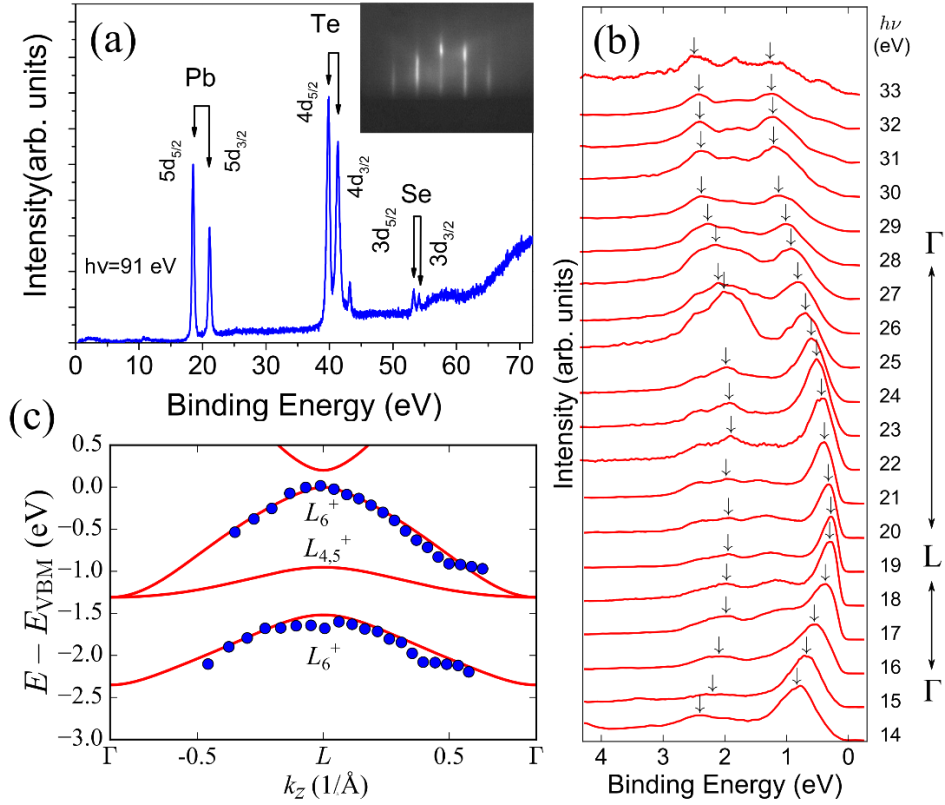

Figure 2. (a) Angle-integrated PES of PbTe using 91.5eV photons. The inset shows the RHEED pattern of a PbTe epitaxial film. (b) Photon energy dependent energy distribution curve (EDC) at normal emission where the arrows show the peaks in EDC. (c) The comparison of the measured peaks with the band dispersion obtained from first-principles calculation along Γ-L-Γ.

In the photoemission process, energy and momentum conservation holds. The momentum of photons is neglectful compared with the momentum of electrons in the spectral range adopted herein. Assuming an unreconstructed surface, the in-plane momentum of electrons is conserved and can be deduced by the relation $k_{\parallel} = \sqrt{2m_e E_{kin}} \sin\theta / \hbar$, where $E_{kin}$ is the measured kinetic energy of outgoing electrons and $\theta$ is the polar emission angle. However, the momentum component perpendicular to the surface is not conserved due to the break of translational symmetry by the surface. The strong dispersion with photon energy indicates that direct (dipole) transition plays an important role in the photoemission of PbTe (111) surface. By varying photon energy we can probe different $k_z$ values. Within the direct transition model of photoemission process, the dependence of $k_z$ on photon energy can be modeled by assuming a free-electron final state[36]:

$$\frac{\hbar^2 k_z^2}{2m_e} - V_0 = E_{kin} \cos^2 \theta \tag{1}$$

where $V_0$ is the inner potential. The valence band maximum at 19 eV is the L point in the second BZ (Figure 2b). The inner potential $V_0$ is estimated to be 5.9 eV, and the $\Gamma$ point in the third BZ is fitted to be corresponding to $h\nu$=41 eV.

## C. The $k_\parallel$ Measurement

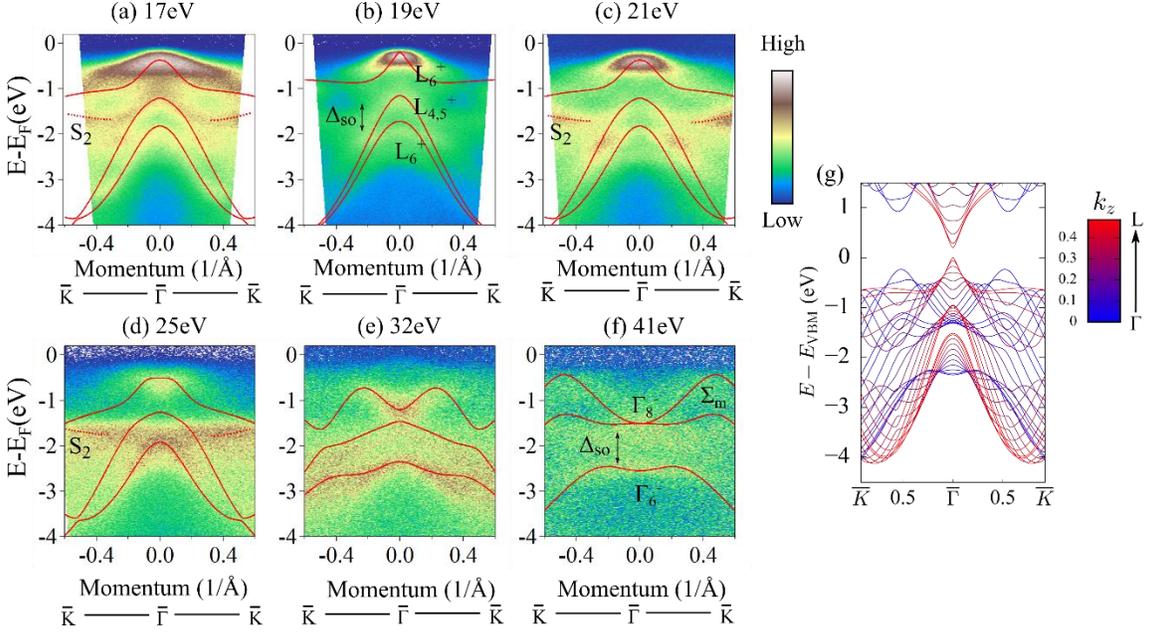

Figure 3. (a-f) ARPES dispersions along $\bar{\Gamma}$ - $\bar{K}$ of a PbTe (111) film taken by horizontal polarized photons with different energies. The lines overlaid are calculated energy band dispersions. Surface states are labeled by dotted lines. (g) The calculated band structure of PbTe (111) projected onto the SBZ with different $k_z$ values shown by the color of the lines.

Figure 3 shows the mapping along the $\bar{\Gamma}$ - $\bar{K}$ cut. Only valence bands are accessible in all mappings, because unintentionally doped PbTe is typically of p-type conductivity due to Pb vacancies. The dispersion is quite different from the results of (100) surface due to the different k-point sampling from the different crystal direction [18, 20]. We find that the overall band dispersion agrees well between the ARPES experiment and the direct transition model calculation in Figure 3(a-f). Figure 3(g) shows the projected band structure of PbTe (111) with different $k_z$, i.e., the calculated band structure with a uniform interval of $k_z$ within the first BZ. Different line colors in Fig. 3(g) indicate varied $k_z$ values. The bulk band structure along the wave vector path $\Gamma$-L, denoted as $k_z$, is projected onto the path connecting $\bar{\Gamma}$ and $\bar{K}$ in the hexagonal SBZ. The resulting counts of energy levels in unit energy interval at each $k_z$ point shown in Figure 3(g) are proportional to the density of states and should resemble the ARPES maps.

The band dispersions passing through $\Sigma_m$ point are shown in Figure 3(f). The $\Sigma$-line valence band maximum ($\Sigma_m$) of PbTe is the second highest valence band. $\Sigma_m$ locates near the midpoint of the $\Sigma$-line in the FCC BZ (Fig.1(a)) and has twelve-fold degeneracy. The convergence of L and $\Sigma_m$ is the key factor to the enhancement of thermoelectric figure of merit of PbTe-based alloy[8]. The energy difference between the L and $\Sigma_m$ point has not been clearly determined in previous studies. By performing ARPES measurements on the (111) surface we can find the one of the $\Sigma_m$ points in the $k_z$ plane across $\Gamma$ point along the $\overline{\Gamma}$ - $\overline{K}$ cut. The energy difference between the L and $\Sigma_m$ point is 0.2 eV.

The spin-orbit coupling plays an important role in PbTe energy bands as the component elements involved are heavy. The value of spin-orbit splitting varies at L and $\Gamma$ point which arises from the different weights of atomic orbital mixing. At L and $\Gamma$ point shown in Figure 3 (b) and (f), the splitting between the second and third valence bands due to spin-orbit interaction is $\Delta_{SO}(L) = E(L_{4,5}^+ - L_6^+) = 0.62$ eV and $\Delta_{SO}(\Gamma) = E(\Gamma_8^- - \Gamma_6^-) = 0.88$ eV, which agrees well with the calculated value of 0.57 eV and 1.05 eV, respectively. These values are approximately three times of 0.2 eV and 0.34 eV at L and $\Gamma$ point in GaAs, respectively [37].

We also note a feature around 1.7 eV below Fermi level in Figure 3(a, c, and d), which does not disperse with photon energy. This feature highlighted by dotted lines cannot fit into the calculated bands by either the direct or the indirect transition model. We assign this band to a surface state (or resonance, $S_2$). The $S_2$ band dominates the photoemission process at h$\nu$=25 eV (Fig. 3(d)), whereas its intensity is much weaker than the direct transition peaks from 14 eV to 23 eV. The intensity variation of $S_2$ in Figure 3 as a function of photon energy can be attributed to matrix element effect[38]. The theoretical calculation could help us to disentangle surface states from the bulk band structures. We further employed a symmetric slab composed of 59 atomic layers (~110 Å) with Te atoms at both ends of PbTe to simulate the surface states. In Figure 4(a), the red dots show the bands with significant weight from surface Te atoms. The surface states are labeled by $S_1$, $S_2$ and $S_3$. The calculated $S_2$ band should be shifted downward about 1.0 eV to match with the experimental data due to the usage of semilocal functional, which usually has an error estimation of band gaps. Though the $S_2$ state is observed only away from $\overline{\Gamma}$ point it could continue through $\overline{\Gamma}$ point or overlap with the bulk bands as seen in the calculated results of Figure 4(a). The difficulty to distinguish the $S_2$ and $L_{4,5}^+$ states near $\bar{\Gamma}$ point is due to their proximity in energy and similarity of band profiles.

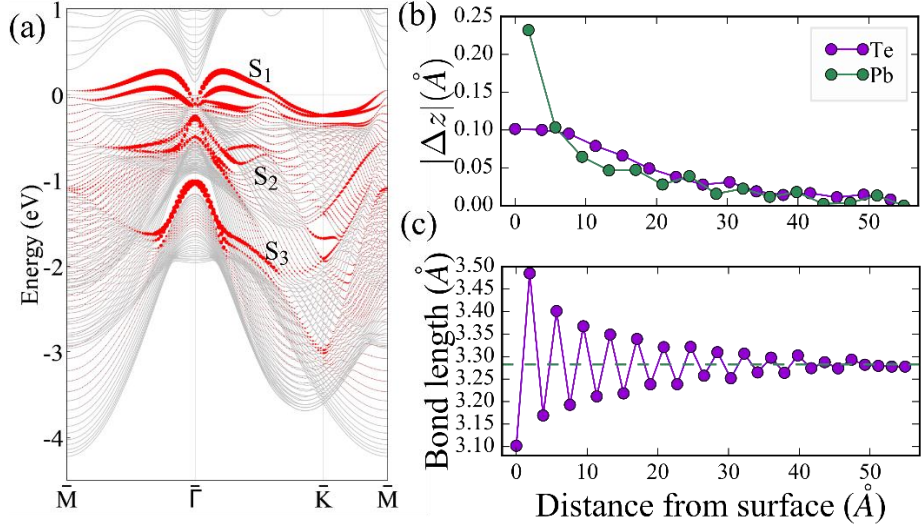

Figure 4. (a) Calculated band structure using a semilocal functional of a PbTe symmetric slab. The size of the red dots marks the weight of wave function projected to the spheres around the first two surface Te atoms. (b) The shift of atomic positions in z direction relative to the bulk positions. (c) The bond length as function of distance from the surface. The positions of the bonds are defined as the positions of starting atoms, and the dashed line in (c) shows the bulk bond length.

We note the hole effective mass discrepancy between the measured ARPES data and the calculated bands, i.e. the hole effective mass of the calculated bands is obviously smaller than that measured by ARPES in the range from 17 eV to 21 eV (Fig. 3 (a-c)). This issue could be explained by the surface states as well. We notice from the calculation result that there are surface states near the valence band edge ($S_1$) which explains the unexpected large effective mass of the valence states. Such surface states are not unfamiliar, e.g. α−GeTe was recently known to show similar surface states[39]. The surface states near the band edge could influence transport and optical properties of PbTe. The effect of surface passivation of PbTe surface states in heterojunctions and nanostructures is not fully explored yet.

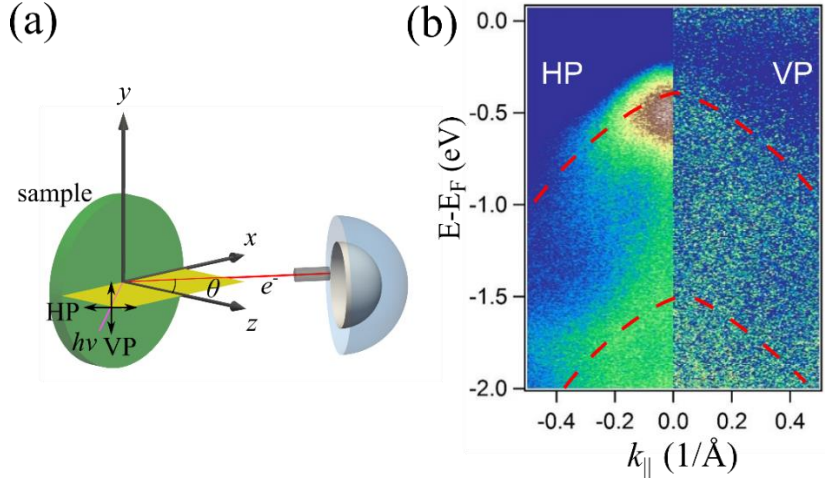

Figure 5. (a) Geometry of the photoemission process. The PbTe sample and the plane defined by incident light and analyzer slit are shown by green and yellow color, respectively. Horizontal and vertical linearly polarized light are marked. (b) APRES mapping along $\bar{\Gamma} - \bar{K}$ by excitation of HP and VP photons.

The orbital character can also be resolved when photons with various polarizations are used in ARPES (Fig. 5(a)). Since the vector potential for vertical polarization (VP) lies in-plane $\bar{\Gamma} - \bar{K}$ direction, it only probes electronic states that do not possess even parity along this direction. The valence band arises from Te $5p_z$ and Pb 6s orbitals has approximately even parity along $\bar{\Gamma} - \bar{K}$ direction, thus their intensity should be largely suppressed with VP photons. This is indeed observed experimentally in Figure 5(b). On the other hand, since the vector potential for horizontal polarization (HP) has both in-plane and out-of-plane components, which allows photoexcitation from orbitals with different symmetries, valence bands show up clearly in this case. Our first-principles calculation confirms that the valence band of PbTe mainly composed of Te 5p and Pb 6s orbitals in which Te 5p contributes 62% while Pb 6s has 36%. As PbTe has a cubic crystal structure, $p_z$, $p_x$, and $p_y$ should have same contribution (21%) to the valence band in a bulk crystal[22]. However, in ARPES data only $p_z$ component can be observed. This phenomenon could be due to the breakdown of the bulk cubic symmetry near the PbTe surface. We tested the idea by examining the slab calculation with relaxed atomic positions. The shift of cations and anions and the change of bond length relative to bulk ones were obtained (see Fig. 4(b),(c)). Especially, the bond length varies between 3.10 Å and 3.48 Å and shows a dramatic change relative to the bulk bond length. The bond lengths gradually restore to the bulk value about 20 Å from the surface. Based on the universal mean-free-path for photoelectrons in solids[36], the electron escape path is estimated to be less than 5 Å for the photon energy adopted in our experiment, corresponding to the first two or three layers in PbTe. Consequently, surface effect is prominent in our photoemission experiment.

The layered distortion near the surface stems from the structure instability of PbTe[33], and is very similar to another group-IV chalcogenides SnSe, with a layer-distorted rock-salt structure, which recently shows a thermoelectric figure of merit of record high along the in-plane *b*-axis of its layered structure[40, 41]. Here, the distorted structure at PbTe (111) surface could be a potential pathway to improve the thermoelectric figure of merit of PbTe.

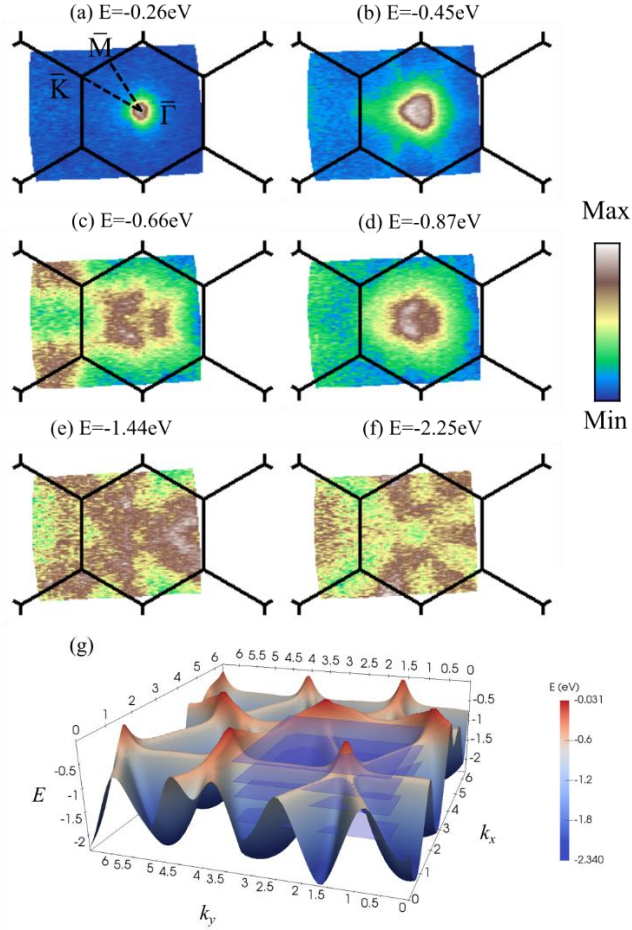

Figure 6. (a-f) A stack of ARPES iso-energetic contours at different energy below Fermi level. The black lines mark the Brillouin zone boundaries. (g) Calculated energy eigenvalues of the first valence band.

We further examine the angular variation of the photoemission spectroscopy. Figure 6 shows the energy contours at several representative energies below the Fermi level. Close to the valence band maximum, the contours are quite isotropic with circular shape, but they become much more anisotropic as the energy decreases, reflecting the underlying bulk band nature as a result of crystal field. In particular, a snowflake shape of spectral weight is observed with six lobes along the $\bar{\Gamma}$ - $\bar{M}$ symmetry line of the hexagonal SBZ between E=-1.44 eV and -2.25 eV, revealing the hexagonal symmetry of the rock-salt structure along [111] direction. These features are well captured in the band structure calculation shown in Figure 6 (g) in which the calculated eigenvalue of the top valence band is given. The hexagonal warping effect leads to a variety of interesting results such as out-of-plane spin polarization for the Dirac cone of the topological insulator $Bi_2Te_3$ [42]. The observed six lobes along $\bar{\Gamma}$ - $\bar{M}$ make large contribution to the electron density of states near the valence band maximum, which is important for the outstanding thermoelectric performance of PbTe.

## IV. Conclusions:

The energy band dispersions of both perpendicular and parallel to the Γ-L direction were obtained by varying the incident photon energies and polarizations in ARPES measurements. The combination of ARPES and first-principles calculation clearly reveal the full 3D band structure, orbital characters, spin-orbit splitting energies, and surface states. The photon energy dependent spectra show the bulk character. Both of L and Σ valence bands are observed and their energy difference is determined to be 0.2 eV. The spin-orbit splitting energies at L and Γ points as large as 0.62 eV and 0.88 eV are identified, respectively. The surface states below and close to the valence band maximum are unambiguously identified. By varying the photon polarization, the valence bands are observed to be composed of a mixture of Pb 6s and Te 5 $p_z$ orbitals with mostly in-plane even parity, which is attributed to the layered distortion in the vicinity of PbTe (111) surface. From the measured constant-energy contours, the valence bands show a transition from isotropic to anisotropic away from L point as a result of crystal field. These findings are of importance not only for fundamental research but also for thermoelectric and optoelectronic device applications based on PbTe.

## V. Acknowledgments

This work was supported by the National Natural Science Foundation of China (Grants Nos. 61290305, 11374259), Thousand Youth Talents Plans, the National R&D program of the MOST of China (Grant No.2016YFA0300203).